\documentclass[prb,amsmath,amssymb]{revtex4}

\newcommand{\bn}[1]{\mbox{\boldmath $#1$}}
\def\e{\mathop{\rm \mbox{{\Large e}}}\nolimits}
\def\p{^{\;\prime}}

\newcommand{\bm}{\mbox{\boldmath $m$}}

\newcommand{\bT}{\mbox{\boldmath $T$}}
\newcommand{\bG}{\mbox{\boldmath $G$}}
\newcommand{\bF}{\mbox{\boldmath $F$}}
\newcommand{\bL}{\mbox{\boldmath $L$}}
\newcommand{\bA}{\mbox{\boldmath $A$}}
\newcommand{\bY}{\mbox{\boldmath $Y$}}
\newcommand{\bW}{\mbox{\boldmath $W$}}
\newcommand{\bB}{\mbox{\boldmath $B$}}
\newcommand{\bP}{\mbox{\boldmath $P$}}
\newcommand{\bM}{\mbox{\boldmath $M$}}
\newcommand{\bI}{\mbox{\boldmath $I$}}
\newcommand{\bZ}{\mbox{\boldmath $Z$}}
\newcommand{\bC}{\mbox{\boldmath $C$}}
\newcommand{\bJ}{\mbox{\boldmath $J$}}
\newcommand{\bD}{\mbox{\boldmath $D$}}

\newcommand{\bcero}{\mbox{\boldmath $0$}}

\newcommand{\bkappa}{\mbox{\boldmath $\kappa$}}
\newcommand{\bPsi}{\mbox{\boldmath $\Psi$}}
\newcommand{\bGamma}{\mbox{\boldmath $\Gamma$}}
\newcommand{\bTheta}{\mbox{\boldmath $\Theta$}}

\newcommand{\Det}{\mbox{{\rm Det}}}

\begin{document}

\title[Transfer Matrices and Green Functions in heterostructures]{Transfer Matrices and Green
Functions for the study of elementary excitations in multilayered heterostructures}

\author{R. P\'erez-\'Alvarez$^{\dag}$\footnote{Permanent address:
Dpto. F\'{\i}sica Te\'orica, Facultad de F\'{\i}sica, Universidad de La Habana, San
L\'azaro y L, Vedado 10400, Cuba. e-mail: rpa@fisica.uh.cu}and F.
Garc\'{\i}a-Moliner$^{\dag}$}

\affiliation{\dag\ Univ. ``Jaume I", Castell\'on de la Plana, Spain.}

\date{\today}

\begin{abstract}
This article is concerned with a mathematical tool, the Associated Transfer
Matrix $\bT$, which proves useful in the study of a wide class of physical problems
involving multilayer heterostructures. General properties of linear, second order
differential matrix Sturm Liouville operators are discussed as a basis for establishing
general properties of $\bT$, which is also generally related to the Green function $\bG$.
Some identities satisfied by $\bT$ are derived, which prove useful in practice to monitor
the numerical quality of computational processes.
\end{abstract}


\maketitle


\section{Introduction}

Problems concerned with multilayer systems arise very often in different fields of
physics. One has to integrate differential equations across some sequence of different
domains, which raises the question of repeated matching at all interfaces involved. In
particular, the development of efficient techniques of epitaxial crystal growth
originated a great deal of work of this nature in solid state physics, notably -but not
only- in the field of semiconductor quantum heterostructures \cite{chino, Vinter91}.

A great deal of activity has also been concerned with {\em Quasiregular Heterostructures}
\cite{RPA01a} which follow some non periodic self-replicative sequence, of which the best
known is the {\em Fibonacci} sequence \cite{Merlin85}. In these cases the number of
interfaces can be very large.

It is therefore of practical value to have mathematical tools available which can be
useful in the study of such systems. Among these, the Surface Green Function Matching
\cite{chino} has been extended to an arbitrarily large number of interfaces
\cite{RPA95b}. Green functions were generally related to a transfer matrix \cite{FGM90}
which transfers amplitudes and normal derivatives -of which more will be said presently-
but it was found later \cite{RPA01c, VVR01a} that it is physically more appealing to
introduce a different kind of transfer matrix, here denoted $\bT$ and termed the {\em
Associated Transfer Matrix} which is even simpler to operate with and is directly related
to the physically interesting magnitudes of the problem under study.

The purpose of this article is to expound on the general relationship between $\bT$ and
the Green function $\bG$, to discuss some basic formal properties of $\bT$ and to derive
from this some general identities satisfied by $\bT$ which can prove useful in practice.
We shall deal explicitly with the case of planar geometry, although the analysis can also
be adapted to other tractable geometries -e.g. spherical or cylindrical.

Irrespective of the specific physical nature of the states, modes, waves or
quasiparticles we may be concerned with, the general setup is the following: We start
from 3D linear, second order differential system and Fourier transform in the 2D $(x,y)$
plane of the interfaces, thus introducing a 2D wavevector $\bkappa$. We then have a
system of $N$ coupled, $\bkappa$-dependent ordinary differential equations in the
variable $z$, in the direction perpendicular to the interfaces. We stress that in
practice $N$ can be sufficiently large that an analytical study is impracticable. For
instance, in electronic band structure calculations one can find multiband envelope
function calculations of electronic band structure with $N=8$ \cite{Szmulowicz2003} and
$N=14$ \cite{Rossler84} and pseudopotential calculations with up to $N=65$
\cite{KitchinThesis}. Thus, it is interesting to study in some detail the formal
properties of an object, $\bT$ \cite{RPA01c}, which is a useful tool for numerical
computation.

The general formulation of the problem is setup in Section 2 and basic questions related
to the hermiticity of linear, second order matrix differential Sturm-Liouville operators
are discussed in Section 3. This provides a basis to study infinite domains and
regularity at infinity -Section 4- after which the general relationship between $\bT$ and
the Green function $\bG$ is established in Section 5. Finally, general properties and
identities satisfied by $\bT$ are derived in Section 6.

\section{Formulation of the problem}

The $N$ coupled differential equations involve $N$ coupled amplitudes corresponding to
the physical model under study -e.g. the components of an electronic wavefunction in an
$N$-band envelope function model or coupled elastic vibration amplitudes and
electrostatic potential in piezoelectric waves. Loosely speaking, we shall refer to the
`primary field', for reasons presently explained.

The $N$ amplitudes of the primary field will be condensed in an object, $\bF(z)$, denoted
for convenience as an `$N$-vector', which need not imply that it transforms literally as
a vector.

We now consider the vast class of physical problems for which the differential system can
be cast in compact matrix form as

{\large
\begin{eqnarray}
\label{Eq01} \bL(z)\cdot\bF(z) = \frac{d\bA(z)}{d z} + \bY(z)\cdot \frac{d\bF(z)}{d z} +
\bW(z) \cdot \bF(z) = \bcero \;,
\end{eqnarray}
}

\noindent where we have defined another $N$-vector, namely

{\large
\begin{eqnarray}
\label{Eq02} \bA(z) = \bB(z) \cdot \frac{d\bF(z)}{dz} + \bP(z) \cdot \bF(z) \;.
\end{eqnarray}
}

We denote this linear differential form, derived from $\bF$, as the {\em secondary
field}. For instance, if $\bF$ describes elastic strains, then $\bA$ describes stresses.
More precisely, normal stress components \cite{chino}. Furthermore, we are concerned with
eigenvalue problems. The eigenvalue parameter, henceforth denoted $\Omega$, is contained
in $\bW$. The dependence on $\bkappa$ and $\Omega$ is understood throughout.

The $z$ dependence of some, or all, of the (matrix) coefficients of (\ref{Eq01}) and
(\ref{Eq02}) is an essential feature of heterostructures, even if piecewise homogeneous
domains are involved.

Now, $\bF(z)$ is obviously continuous everywhere and, furthermore, integration across any
of the interfaces of the heterostructure proves that $\bA(z)$ is also continuous, which
bears out the convenience of this compact formulation. Since both, the primary field and
the secondary field are continuous, this suggests the convenience of defining, instead of
a transfer matrix $\bM$ \cite{chino,FGM90} which transfers $\bF(z)$ and $\bF\p(z)$, a
different one, here denoted as the Associated Transfer Matrix (ATM) $\bT$, which
transfers the two fields that are continuous everywhere. Thus we define the $2N$- vector

{\large
\begin{eqnarray}
\label{Eq03}
\bPsi(z) = \begin{array}{|c|}
           \bF(z) \\
           \bA(z)
           \end{array}
\end{eqnarray}
}

\noindent and then, starting from any initial point $z_0$, we define $\bT(z,z_0)$ which
transfers $\bPsi$ from $z_0$ to any $z$ by

{\large
\begin{eqnarray}
\label{Eq04} \bPsi(z) = \bT(z,z_0) \cdot \bPsi(z_0) \;.
\end{eqnarray}
}

Unlike the {\em coefficient transfer matrix}, often used for systems with piecewise
constant coefficients \cite{Erdos82,PPP95}, $\bM(z,z_0)$ and $\bT(z,z_0)$ do not depend
on knowing a basis of $2N$ linearly independent solutions. Both, $\bM(z,z_0)$ and
$\bT(z,z_0)$, pertain to initial value problems, while the Green function $\bG(z,z\p)$
depends on the choice of boundary conditions. Thus, for a given differential system $\bM$
and $\bT$ are unique, while $\bG$ is not. The relationship between $\bG$ and $\bM$ was
discussed in \cite{chino,FGM90,RPA88c}. Here we discuss the relationship between $\bG$ and
$\bT$, for which we first study the general properties of the latter, for which first
some observations concerning the differential system are in order.

\section{The Hermiticity of the differential operator}

The differential matrix $\bL(z)$ of (\ref{Eq01}) is the representation of a linear
differential operator which we now discuss. We recall \cite{Friedman} that the full
definition of the operator requires the specification of the manifold ${\cal S}$ of
functions -here termed `vectors' $\bF(z)$- on which it acts.

Let $\hat{p}$ be an operator represented by the operating rule

{\large
\begin{eqnarray}
\label{Eq05} p = - i \frac{d}{dz} \;.
\end{eqnarray}
}

\noindent So far, $\hat{p}$ is formally Hermitean \cite{Friedman}, [Chapter 3]. Now let
$\hat{\bL}$, represented by the operating rule (\ref{Eq01}), be an operator defined to
act on a given manifold ${\cal S}$ and let $\bF$ denote the generic function of this
particular manifold. We write the system (\ref{Eq01}) explicitly as

{\large
\begin{eqnarray}
\label{Eq06} \bL(z) \cdot \bF(z) &=&  \frac{d}{dz}\left[\bB(z) \cdot \frac{d\bF(z)}{dz} +
\bP(z) \cdot \bF(z) \right] + \nonumber \\ & & \bY(z) \cdot \frac{d\bF(z)}{dz} +\bW(z)
\cdot \bF(z) \;.
\end{eqnarray}
}

Thus, the operator $\hat{\bL}$ represented by this equation is

{\large
\begin{eqnarray}
\label{Eq07} \hat{\bL} &=&  -\hat{p}\left[\bB \cdot \hat{p} + i\bP \cdot \right]
-i\bY\cdot \hat{p} +\bW(z) \cdot \,,
\end{eqnarray}
}

\noindent where the dots indicate that, as a matrix, this is to act on vectors $\bF$.
Furthermore, in most physical problems related to heterostructures the matrix
coefficients satisfy the conditions

{\large
\begin{eqnarray}
\label{Eq08} \bB = \bB^\dag \;,\; \bP = - \bY^\dag \;,\; \bW = \bW^\dag \;.
\end{eqnarray}
}

Thus, for the large class of problems we are concerned with we assume (\ref{Eq08}) to
hold. Then it is easily seen that the {\em Hermitean adjoint} of $\hat{\bL}$ is

{\large
\begin{eqnarray}
\label{Eq09} \hat{\bL}^\dag &=&  -\hat{p}^\dag \bB^\dag \cdot \hat{p}^\dag + i
\hat{p}^\dag \bY^\dag \cdot +i\bP^\dag \cdot \hat{p}^\dag +\bW^\dag \cdot  \;,
\end{eqnarray}
}

\noindent which so far only ensures that $\hat{\bL}$ is {\em formally Hermitean}.

We now, more specifically, define ${\cal S}$ as the manifold of functions defined in a
given interval $[a,b]$ -provisionally assumed to be finite- and satisfying given boundary
conditions at its extremes. A physical example could be a multiple quantum well with
infinite barriers -vanishing amplitudes- in the external media. Let us then consider
another operator $\hat{\bL}_2$, acting on another manifold ${\cal S}_2$ of functions
$\bF_2$ defined in the same interval $[a,b]$ but otherwise satisfying so far unspecified
boundary conditions.

We define $\hat{\bL}_2$ to be represented by

{\large
\begin{eqnarray}
\label{Eq10} \bL_2(z) \cdot \bF_2(z) &=&  \frac{d}{dz}\left[\bB^\dag(z) \cdot
\frac{d\bF_2(z)}{dz} - \bY^\dag(z) \cdot \bF_2(z) \right] - \nonumber \\ & & \bP^\dag(z)
\cdot \frac{d\bF_2(z)}{dz} +\bW^\dag(z) \cdot \bF_2(z)
\end{eqnarray}
}

\noindent and then at every point $z$ of the interval $[a,b]$, by assumption common to
${\cal S}$ and ${\cal S}_2$, we define the {\em residual} ${\cal R}(z)$ by

{\large
\begin{eqnarray}
\label{Eq11} {\cal R}(z) &=& \left[ \frac{d\bF^\dag(z)}{dz} \cdot \bB^\dag(z) \cdot
\bF_2(z) - \bF^\dag(z) \cdot \bB^\dag(z) \cdot \frac{\bF_2(z)}{dz} \right. \nonumber \\ &
&  \left. + \bF^\dag(z)\cdot\bP^\dag(z)\cdot\bF_2(z)
+\bF^\dag(z)\cdot\bY^\dag(z)\cdot\bF_2(z) \right]^\dag \;.
\end{eqnarray}
}

We then study the integral

{\large
\begin{eqnarray}
\label{Eq12} \langle \bF_2 \vert \hat{\bL} \bF \rangle = \int_a^b dz \; \bF_2^\dag(z)
\cdot \bL(z) \cdot \bF(z)
\end{eqnarray}
}

\noindent which, after partial integration, is

{\large
\begin{eqnarray}
\label{Eq13} \langle \bF_2 \vert \hat{\bL} \bF \rangle = \langle \bF \vert \hat{\bL}_2
\bF_2 \rangle^\dag + {\cal R}(b) - {\cal R}(a) \;.
\end{eqnarray}
}

Now, since the matrix coefficients satisfy the conditions (\ref{Eq08}), the operating
rule for $\hat{\bL}_2$ is the same as for $\hat{\bL}$, whence follows that the residual
is

{\large
\begin{eqnarray}
\label{Eq14} {\cal R}(z) = \bF_2^\dag(z)\cdot\bA(z)-\bA_2^\dag(z)\cdot \bF(z)
\end{eqnarray}
}

\noindent and thus the residual difference appearing in (\ref{Eq13}) is

{\large
\begin{eqnarray}
\label{Eq15} {\cal R}(b) - {\cal R}(a) &=&
\bF_2^\dag(b)\cdot\bA(b)-\bA_2^\dag(b)\cdot\bF(b) \nonumber \\
& & -\bF_2^\dag(a)\cdot\bA(a)+\bA_2^\dag(a)\cdot\bF(a) \;.
\end{eqnarray}
}

\noindent The mathematical definition of ${\cal R}$ acquires a physical meaning in terms
of the primary and secondary fields, the two objects transferred by $\bT$.

This is the central point of the analysis. The operators $\hat{\bL}$ and $\hat{\bL}_2$
are not yet fully defined, as the boundary conditions have not been specified. If we
assume {\em only} that ${\cal S}$ and ${\cal S}_2$ are such that the combined effect of
their boundary conditions is such that ${\cal R}(b)-{\cal R}(a)$ vanishes, then

{\large
\begin{eqnarray}
\label{Eq16} \langle \bF_2 \vert \hat{\bL} \bF \rangle = \langle \bF \vert \hat{\bL}_2
\bF_2 \rangle^\dag
\end{eqnarray}
}

\noindent and $\hat{\bL}_2$ is the Hermitean adjoint of $\hat{\bL}$.

If, furthermore, the boundary conditions are the same separately in ${\cal S}$ and ${\cal
S}_2$, then $\hat{\bL}_2$ -the Hermitean adjoint of $\hat{\bL}$- is also equal to
$\hat{\bL}$ and it is then that $\hat{\bL}$ is really {\em Hermitean}, that is to say,
technically, it is {\em totally Hermitean} \cite{Friedman}. Such is the case, for
instance, for electronic states with external infinite barriers, where all amplitudes
vanish, or for elastic waves with external free surfaces, where all normal stresses
vanish.

It seems reasonable to put forward this type of analysis to justify the basis of standard
calculations for physical models with $N>1$. Furthermore, this is the starting point to
analyse a question directly relevant to the physical models employed in practice to study
heterostructures.

\section{Infinite domains: Causality and regularity at infinity}

This situation arises if the external media are semiinfinite or, simply, if one studies a
bulk medium. So far it has been implicitly assumed that the eigenvalues are real, which
is formally correct if the interval $[a,b]$ is finite. We must now analyse the situation
when this is infinite, for which a brief reminder of some basic facts is at this stage
convenient. In order to fix ideas it suffices to consider a simple case with $N=1$.

Firstly, we must define a sign convention for the description of stationary amplitudes
$F(z,t)$. We adopt the convention

{\large
\begin{eqnarray}
\label{Eq17} F(z,t) = F_0 \; \e^{i(kz-\omega t)}
\end{eqnarray}
}

\noindent for a wave travelling to the right. Then, when in the time dependent picture
$t\rightarrow\infty$, in the stationary state time independent amplitude,

{\large
\begin{eqnarray}
\label{Eq18} F(z) = F_0 \; \e^{ikz} \;,
\end{eqnarray}
}

\noindent the variable $z\rightarrow\infty$. In order for this to be a physically
acceptable regular solution, it cannot blow up for $z\rightarrow\infty$, which requires
formally (i) that the real wavevector $k$, corresponding to the allowed eigenvalue of a
stationary propagating eigenstate, be defined as

{\large
\begin{eqnarray}
\label{Eq19} k = \lim_{\epsilon\rightarrow 0} (k+i\epsilon)
\end{eqnarray}
}

\noindent and (ii) that we define the limit of $F(z)$ for $z\rightarrow\infty$ as the
{\em regular limit}.

{\large
\begin{eqnarray}
\label{Eq20} \lim_{\epsilon\rightarrow 0} \lim_{z\rightarrow\infty} F_0
\e^{i(k+i\epsilon)z} &=& 0 \;.
\end{eqnarray}
}

\noindent The same analysis holds obviously for waves travelling to the left and
$z\rightarrow-\infty$ and the same basic facts hold for less simple cases, e.g. for any
$N>1$.

Let us now return to the standard operator $\hat{\bL}$ on which the analysis of the class
of problems here considered is based. Since this is defined to act on the manifold ${\cal
S}$ of functions which are regular at infinity, the second and fourth terms of the
residual difference (\ref{Eq15}) vanish. However in order for (\ref{Eq15}) to vanish
entirely, the condition on the adjoint of $\hat{\bL}$ is that the boundary conditions of
regularity at infinity must hold for the $\bF_2^\dag$, not the $\bF_2$. In fact, in order
for (\ref{Eq15}) to vanish the $\bF_2$ must be precisely irregular at infinity, as only
then the $\bF_2^\dag$ are regular. Therefore the manifold ${\cal S}_2$ is not the same as
${\cal S}$ and we reach the non trivial conclusion that $\hat{\bL}$, the operator
normally used in the description of infinite -or, in the event- seminfinite regular media
is {\em not totally Hermitean}.

Although a full analysis in general terms (\cite{UJILibro}, \S 7.5) of the conditions at
infinity is outside the scope of this article, it is in order to recall that while
regularity at infinity is the physical requirement in the stationary state -time
independent- picture, causality is the condition required in standard physical theory,
and this in turn requires $\omega$ to be defined as the limit of $(\omega +i\eta)$ for
$\eta\rightarrow 0$. Now, for the class of problems here considered $\omega$ bears a
direct relation to the eigenvalue variable $\Omega$. For instance, for electronic states
$\Omega$ is the energy and $\omega$ is $\Omega/\hbar$; for most wave problems $\omega$ is
$\Omega^{1/2}$, etc. Then the causal formulation amounts to requiring that $\Omega$ is to
be read everywhere as

{\large
\begin{eqnarray}
\label{Eq21} \Omega = \lim_{\eta\rightarrow 0} (\Omega + i\eta) \;.
\end{eqnarray}
}

Even for bound states, for which the eigenvalue is real, a small imaginary part may have
to be added in practice, for instance, when calculating spectral functions numerically
with a computer code. The point to stress is that (\ref{Eq21}) must be kept as such
everywhere and this has a direct bearing on the foregoing analysis.

It is correct to define $\hat{\bL}^\dag$, the {\em Hermitean conjugate} of an {\em
operator} $\hat{\bL}$ as has been done, but in the matrices resulting in the
$z$-representation, where $\Omega$ appears explicitly, it must always be kept
$\Omega+i\eta$, whether $\eta$ stays small but finite or it tends to zero as in
(\ref{Eq21}). In other words, while the standard {\em Hermitean conjugate} of a {\em
matrix} of the general form

{\large
\begin{eqnarray}
\label{Eq22} \bm(\Omega) = \bm_1(\Omega) + i \bm_2(\Omega)
\end{eqnarray}
}

\noindent is

{\large
\begin{eqnarray}
\label{Eq23} \bm^\dag(\Omega) = \bm_1^t(\Omega^*) - i \bm_2^t(\Omega^*) \;,
\end{eqnarray}
}

\noindent this is not the concept required for an explicit matrix analysis of the
problems here considered. The concept of {\em transconjugate} matrix

{\large
\begin{eqnarray}
\label{Eq24} \bm^c(\Omega) = \bm_1^t(\Omega) - i \bm_2^t(\Omega)
\end{eqnarray}
}

\noindent was used in \cite{FGM90} where the relationship between the standard transfer
matrix $\bM$ and the Green function $\bG$ was established. This was further developed in
\cite{chino}. If one used $\bm^\dag$, instead of $\bm^c$, then one would obtain a
negative density of states. We now relate $\bG$ to $\bT$, defined in (\ref{Eq04}), while
extending significantly the analysis.

\section{The relationship between $\bG$ and $\bT$}

We now start from

{\large
\begin{eqnarray}
\label{Eq25} \bL(z) \cdot \bG(z,z\p) = \bI_N \; \delta(z-z\p) \;,
\end{eqnarray}
}

\noindent where $\bI_N$ is the $N\times N$ unit matrix and $\bL(z)$ is the differential
matrix of (\ref{Eq01}) but, instead of $\bA(z)$ of (\ref{Eq02}), the differential form is
now

{\large
\begin{eqnarray}
\label{Eq26} \bA(z,z\p) &=& \bB(z)\cdot\frac{\partial\bG(z,z\p)}{\partial z} + \bP(z)
\cdot \bG(z,z\p) \;.
\end{eqnarray}
}

Defining

{\large
\begin{eqnarray}
\label{Eq27} {\bn {\cal A}}^\pm(z) &=& \lim_{z\p\rightarrow z\pm0} \bA(z,z\p) \;,
\end{eqnarray}
}

\noindent a first integration of (\ref{Eq25}) yields the identity

{\large
\begin{eqnarray}
\label{Eq28} {\bn {\cal A}}^+(z) - {\bn {\cal A}}^-(z) &=& -\bI_N \;.
\end{eqnarray}
}

Furthermore, a detailed formal analysis of (\ref{Eq25}) shows \cite{UJILibro} that if we
define

{\large
\begin{eqnarray}
\label{Eq29} \bZ(z,z\p) &=& \frac{\partial \bG(z,z\p)}{\partial z\p} \cdot \bB^c(z\p) +
\bG(z,z\p) \cdot \bP^c(z\p)
\end{eqnarray}
}

\noindent and hence

{\large
\begin{eqnarray}
\label{Eq30} ^\pm{\bn {\cal Z}}(z) &=& \lim_{z\p\rightarrow z\mp 0} \bZ(z,z\p) \;,
\end{eqnarray}
}

\noindent then there is another identity, namely

{\large
\begin{eqnarray}
\label{Eq31} ^+{\bn {\cal Z}}(z) - ^-{\bn {\cal Z}}(z) &=& -\bI_N \;.
\end{eqnarray}
}

It is now convenient to display $\bT$ in the form

{\large
\begin{eqnarray}
\label{Eq32} \bT &=& \begin{array}{||cc||}
                     \bT_{AA} & \bT_{AD} \\
                     \bT_{DA} & \bT_{DD}
                     \end{array} \;,
\end{eqnarray}
}

\noindent where the submatrices $\bT_{\alpha\beta}$ ($\alpha,\beta=A,D$) are $N\times N$.
Then, similarly to (\ref{Eq03}), we define the rectangular $2N\times N$ matrix

{\large
\begin{eqnarray}
\label{Eq33} \bGamma(z,z\p) = \begin{array}{|c|}
           \bG(z,z\p) \\
           \bA(z,z\p)
           \end{array}
\end{eqnarray}
}

\noindent and thus

{\large
\begin{eqnarray}
\label{Eq34} \bGamma(z,z\p) = \bT(z,z_0)\cdot\bGamma(z_0,z\p)\;,
\end{eqnarray}
}

\noindent that is to say, for fixed $z\p$ $\bGamma(z,z\p)$ is transferred like $\bPsi(z)$
and, in particular

{\large
\begin{eqnarray}
\label{Eq35} \bG(z,z\p) = \bT_{AA}(z,z_0)\cdot\bG(z_0,z\p) +
\bT_{AD}(z,z_0)\cdot\bA(z_0,z\p)\;. \nonumber  \\
\end{eqnarray}
}

A similar analysis can be carried out with the Green function of the Hermitean conjugate
differential system, which has the same ATM. This leads to a formula analogous to
(\ref{Eq35}) in which $\bT^c(z\p,z_0)$, acting from the right, transfers in $z\p$ with
$z$ fixed.

Following this line of argument and recalling that the definition of $\bT$ does not
depend on the choice of $z_0$, we arrive -details in \cite{UJILibro}- at a general
expression for $\bG$ in terms of $\bT$ of the form

{\large
\begin{eqnarray}
\label{Eq36} \bG(z,z\p) = \left\{
\begin{array}{lr}
\sum_{\alpha\beta} \bT_{A\alpha}(z,z_0) \cdot \bC^<_{\alpha\beta}
                   \cdot \bT^c_{A\beta}(z\p,z_0) & z \leq z\p \\
&  \\ \sum_{\alpha\beta} \bT_{A\alpha}(z,z_0) \cdot \bC^>_{\alpha\beta}
                   \cdot \bT^c_{A\beta}(z\p,z_0) & z \geq z\p\;,
\end{array}
\right.
\end{eqnarray}
}

\noindent for any pair of values of $(z,z\p)$ irrespective of the position of $z_0$.
Hence

{\large
\begin{eqnarray}
\label{Eq37} \bA(z,z\p) = \left\{
\begin{array}{lr}
\sum_{\alpha\beta} \bT_{D\alpha}(z,z_0) \cdot \bC^<_{\alpha\beta}
                   \cdot \bT^c_{A\beta}(z\p,z_0) & z \leq z\p \\
&  \\
\sum_{\alpha\beta} \bT_{D\alpha}(z,z_0) \cdot \bC^>_{\alpha\beta}
                   \cdot \bT^c_{A\beta}(z\p,z_0) & z \geq z\p \;.
\end{array}.
\right.
\end{eqnarray}
}

Not all the eight (matrix) coefficients $\bC_{\alpha\beta}^>$, $\bC_{\alpha\beta}^<$ are
independent. Firstly, whether we proceed from the first or the second row of
(\ref{Eq36}), we must obtain

{\large
\begin{eqnarray}
\label{Eq38} {\bn {\cal G}}(z) \equiv \bG(z,z) = \bC_{AA}^<=\bC_{AA}^>\equiv \bC_{AA} \;.
\end{eqnarray}
}

Then, from (\ref{Eq28}) and (\ref{Eq37}):

{\large
\begin{eqnarray}
\label{Eq39} \bC_{DA}^<-\bC_{DA}^> &=& -\bI_N \;.
\end{eqnarray}
}

Furthermore, an expression for $\bZ(z,z\p)$ (\ref{Eq29}) can be obtained similar to
(\ref{Eq37}) and then, from (\ref{Eq31})

{\large
\begin{eqnarray}
\label{Eq39a} \bC_{AD}^>-\bC_{AD}^< &=& -\bI_N \;.
\end{eqnarray}
}

Now, consider the configuration $z\le z_0 \le z\p$. Then \cite{chino}

{\large
\begin{eqnarray}
\label{Eq40} \bG(z,z\p) &=& \bG(z,z_0)\cdot\left[{\bn {\cal G}}(z_0)\right]^{-1} \cdot
\bG(z_0,z\p) \;.
\end{eqnarray}
}

From this and the above equalities -together with the similar expressions for
$\bZ(z,z\p)$- we obtain

{\large
\begin{eqnarray}
\label{Eq41} \bC_{DD}^< = \bC_{DA}^<\cdot\bC_{AA}^{-1}\cdot\bC_{AD}^< \;; \nonumber \\
\label{Eq42} \bC_{DD}^> = \bC_{DA}^>\cdot\bC_{AA}^{-1}\cdot\bC_{AD}^> \;.
\end{eqnarray}
}

These relationships hold quite generally and reduce the number of independent
coefficients. The rest of the analysis depends on the domain under study.

For the finite internal domains of a heterostructure, it was shown in \cite{RPA88c} that
one can define any arbitrary extended pseudomedium with a Green function satisfying
arbitrary boundary conditions -except for infinite barriers- at the extremes of the said
domains. In practice a considerable simplification of the algebra can be achieved in this
way. The external domains are different. Sometimes they are terminated by infinite
barriers, which is part of the model. The algebra is then fairly simple. A different
situation arises if the external domains are assumed to extend to infinity. We now
consider an infinite regular medium, which is in any case always a valid description for
any domain, even if it is finite or semiinfinite.

We take $z_0=0$ and abbreviate $\bT_{\alpha\beta}(z,z_0)$/$\bT_{\alpha\beta}(z\p,z_0)$ as
$\bT_{\alpha\beta}(z)$/$\bT_{\alpha\beta}(z\p)$. Consider the first row of (\ref{Eq36})
and take $z\p=0$. Then

{\large
\begin{eqnarray}
\label{Eq44} \bG(z,0) = \bT_{AD}(z) \cdot \left\{ \left[\bT_{AD}(z)\right]^{-1}
\dot\bT_{AA}(z) \cdot \bC_{AA} +\bC_{DA}^< \right\} \;.
\end{eqnarray}
}

Regularity at $-\infty$ requires the vanishing of the {\em regular limit} of
(\ref{Eq44}). To this effect we note that even if the coefficients of the differential
system are variable, for sufficient large $\vert z \vert$ these can be asymptotically
`flattened' -i.e. replaced by their average values- at sufficiently large distances
without essential loss of accuracy. Then $\bT$ can be written as $\bT(\epsilon,z)$ and
regular limits taken in the manner of (\ref{Eq19})-(\ref{Eq20}), so we define the regular
limits

{\large
\begin{eqnarray}
\label{Eq45} \bT_\pm &=& \lim_{\epsilon\rightarrow 0} \lim_{z\rightarrow\pm\infty}
\left\{ \left[ \bT_{AD}(\epsilon,z)\right]^{-1}\cdot\bT_{AA}(\epsilon,z) \right\}
\nonumber \\
\label{Eq46} \bTheta_\pm &=& \lim_{\epsilon\rightarrow 0}
\lim_{z\rightarrow\pm\infty} \left\{ \bT_{AA}^c(\epsilon,z) \cdot \left[
\bT_{AD}^c(\epsilon,z)\right]^{-1}\right\}\;.
\end{eqnarray}
}

Now, the first factor of (\ref{Eq44}) has asymptotic oscillatory behaviour. Thus the
regular vanishing of $\bG(-\infty,0)$ requires the vanishing of the second factor in the
regular limit. Following this line of argument and using the general identities just
presented, we obtain the full set of parameters for the regular Green function

{\large
\begin{eqnarray}
\label{Eq47} \bC_{AA} = \left[\bT_--\bT_+\right]^{-1} \;, \nonumber \\ \label{Eq48}
\bC_{DA}^< = -\bT_-\cdot\bC_{AA} \;,\; \bC_{DA}^> = -\bT_+\cdot\bC_{AA} \;, \nonumber \\
\label{Eq49} \bC_{AD}^< = -\bC_{AA}\cdot \bTheta_+ \;,\; \bC_{AD}^> =
-\bC_{AA}\cdot\bTheta_- \;, \nonumber \\ \label{Eq50} \bC_{DD}^< = -\bT_-\cdot\bC_{AA}
\cdot \bTheta_+ \;,\; \bC_{DD}^> = \bT_+\cdot\bC_{AA} \cdot \bTheta_- \;.
\end{eqnarray}
}

The regular Green function is thus fully determined from $\bT$. The latter is always an
ultimate resort if the analytical solution is not viable, while $\bG$ is a natural object
to extract physical information directly from it. As indicated above, the set of
parameters $\bC_{\alpha\beta}^<$, $\bC_{\alpha\beta}^>$ can be considerably simplified by
the convenient choice of simpler boundary condition.

\section{The symplectic character and general properties of $\bT$}

Firstly we note that, due to the fact that the two fields transferred by $\bT$ are
continuous everywhere, on crossing interfaces separating different media $\bT$ is
obtained by simple chain multiplication of the corresponding different matrices. This is
a practical useful feature for the description of heterostructures, not shared by $\bM$
\cite{RPA88c}. Furthermore, let us define the $2N\times 2N$ auxiliary matrix

{\large
\begin{eqnarray}
\label{Eq51} \bJ =&=& \begin{array}{||cc||}
                      \bcero_N & -\bI_N \\
                      \bI_N & \bcero_N
                      \end{array} \;.
\end{eqnarray}
}

Then

{\large
\begin{eqnarray}
\label{Eq52} \bPsi^c\cdot\bJ\cdot\bPsi &=& -\bF^c\cdot\bA+\bA^c\cdot\bF \;.
\end{eqnarray}
}

Now, the general definition of the flux density is

{\large
\begin{eqnarray}
\label{Eq53} j(z) = i\left[\bF^c(z)\cdot\bA(z)-\bA^c(z)\cdot\bF(z)\right] = -i
\bPsi^c(z)\cdot\bJ\cdot\bPsi(z) \;. \nonumber \\
\end{eqnarray}
}

On the other hand, the second order differential system for $\bF$ is transformed in the
standard way into the first order differential system for $\bPsi$,

{\large
\begin{eqnarray}
\label{Eq54} \frac{d\bPsi(z)}{dz} &=& \bD(z)\cdot\bPsi(z)
\end{eqnarray}
}

\noindent with

{\large
\begin{eqnarray}
\label{Eq55} \bD(z) = \begin{array}{||cc||}
             - \bB^{-1}(z) \cdot\bP(z) &
               \bB^{-1}(z) \\
               \left[ \bY(z) \cdot \bB^{-1}(z) \cdot \bP(z) - \bW(z) \right] &
             - \bY(z) \cdot \bB^{-1}(z)
             \end{array} \;. \nonumber \\
\end{eqnarray}
}

It is easily seen that

{\large
\begin{eqnarray}
\label{Eq56} \frac{d}{dz}\left[\bF^c\cdot\bA-\bA^c\cdot\bF\right] = -\bPsi^c \cdot \left[
\bD^c\cdot\bJ+\bJ\cdot\bD \right] \cdot \bPsi \;.
\end{eqnarray}
}

Then, since

{\large
\begin{eqnarray}
\label{Eq57} \bD^c\cdot\bJ+\bJ\cdot\bD  = \bcero_{2N}
\end{eqnarray}
}

\noindent it follows that $j(z)$ is constant everywhere, which is a general expression of
the continuity equation.

Now consider any arbitrary pair of points $(z_0,z)$, transfer $\bPsi$ from $z_0$ to $z$,
write $\bT(z,z_0)$ for brevity as $\bT$ and equate $j(z)$ to $j(z_0)$. This yields

{\large
\begin{eqnarray}
\label{Eq58} \bJ = \bT^c\cdot\bJ\cdot\bT
\end{eqnarray}
}

Thus $\bT$ is a symplectic matrix in a general sense \cite{Arvind}. Since $\bJ$ has unit
determinant, it follows that

{\large
\begin{eqnarray}
\label{Eq59} \left\vert \Det \left[\bT\right]\right\vert^2 = 1
\end{eqnarray}
}

This provides a concise practical rule which can be used to monitor the quality of the
numerical computations as $z$ grows away from $z_0$. Furthermore, by equating the
submatrices of (\ref{Eq58}) we obtain the three independent identities

{\large
\begin{eqnarray}
\label{Eq60} \bT_{AA}^c \cdot \bT_{DA} - \bT_{DA}^c \cdot \bT_{AA} & = & \bcero_N
\\
\label{Eq61} \bT_{DD}^c \cdot \bT_{AD} - \bT_{AD}^c \cdot \bT_{DD} & = & \bcero_N
\\
\label{Eq62} \bT_{AA}^c \cdot \bT_{DD} - \bT_{DA}^c \cdot \bT_{AD} & = & \bI_N \;.
\end{eqnarray}
}

These can also prove useful in two ways, namely: (i) they can help simplify the algebra
in the analytical formulation of a problem and (ii) other concise rules for numerical
monitoring can also be extracted from them.

Further details and different ways to combine $\bT$ and $\bG$ for the study of multilayer
heterostructures are given in \cite{UJILibro}.

\section{Conclusion}

The Associated Transfer Matrix $\bT$ appears to be both, formally appealing and
practically useful. It suggests itself in a natural way from the physics of the problem.
It is unique and does not depend on having an analytical basis, so it can always be an
ultimate resort, obtainable by numerical integration. It satisfies some general
identities, not satisfied by other transfer matrices, which provides useful rules to
monitor the numerical quality of computational processes. Finally, it is directly related
to the Green function, from which direct physical information follows in a natural way.

\underline{Acknowledgements}. This work was done while one of us (R P-A) was enjoying the
hospitality and finantial support of the Universitat `Jaume I' -Castell\'on de la Plana,
Spain- which is here gratefully acknowledged. We also express our appreciation for the
invaluable help of Pilar Jim\'enez in the preparation of the article.

\end{document}